# Crystallographic effects on transgranular chloride-induced stress corrosion crack propagation of arc welded austenitic stainless steel

Haozheng J. Qu[1*], Fei Tao[2], Nianju Gu[2], Timothy Montoya[3], Jason M. Taylor[3], Rebecca F. Schaller[3], Eric Schindelholz[4], Janelle P. Wharry[1]

**Affiliations:**

[1]School of Materials Engineering, Purdue University, IN, USA.

[2]School of Aeronautics and Astronautics, Purdue University, IN, USA.

[3]Center for Materials Science and Engineering, Sandia National Laboratories, NM, USA.

[4]School of Materials Science and Engineering, The Ohio State University, OH, USA.

**\*Correspondence to:**

Haozheng J. Qu,

School of Materials Engineering, Purdue University

**Mail:** 205 Gates Road, Flex Lab, West Lafayette, IN 47906

**Telephone:** +1 (765) 479 - 5157

**Email:** qu34@purdue.edu

Abstract:

The effect of crystallography on transgranular chloride-induced stress corrosion cracking (TGCISCC) of arc welded 304L austenitic stainless steel is studied on >300 grains along crack paths. Schmid and Taylor factor mismatches across grain boundaries (GBs) reveal that cracks propagate either from a hard to soft grain, which can be explained merely by mechanical arguments, or soft to hard grain. In the latter case, finite element analysis reveals that TGCISCC will arrest at GBs without sufficient mechanical stress, favorable crystallographic orientations, or crack tip corrosion. GB type does not play a significant role in determining TGCISCC cracking behavior nor susceptibility. TGCISCC crack behaviors at GBs are discussed in the context of the competition between mechanical, crystallographic, and corrosion factors.

## Introduction

Stress corrosion cracking (SCC) is a longstanding critical materials challenge in austenitic stainless steels. Intergranular SCC (IGSCC) in austenitic steels can largely be controlled by reducing grain boundary sensitization (i.e., Cr depletion) through low-carbon alloying (e.g., 304L), and grain boundary engineering [1], but transgranular SCC (TGSCC) remains an active degradation mode even in low-carbon austenitic steels [2] and their weldments [3], primarily in chloride-based corrosive environments. Chloride-induced stress corrosion cracking (CISCC) is especially problematic because it occurs in standard atmospheric conditions, particularly in coastal environments, over temperatures ranging from ambient to the boiling point of water, and at modest stresses near the proportional limit of solution annealed austenitic stainless steels [4]. CISCC has been reported in austenitic stainless steels across a wide range of applications, including nuclear power reactors [5], petrochemical pipelines and valves [6,7], and rock climbing gear [8]. More recently, there has been mounting concern regarding the potential for CISCC along arc weld seams on austenitic stainless steel, including SS304/304L and SS316/316L, canisters used as spent nuclear fuel (SNF) dry storage containers [9,10]. Welds, especially those produced by high heat-input methods such as arc welding, are even more susceptible to CISCC because of their tensile residual stresses [11,12]. Besides, since these SNF canisters are often stored in coastal regions for natural air circulation, the high humidity and deliquescent chloride-rich sea salt in the moist coastal air provides the perfect ingredient for CISCC [5,13,14].

TGSCC cracks tend to propagate from pits in a direction perpendicular to the principal tensile stress [15–17]. The extension of a pit into a microcrack is associated with the maximum principal strain, which occurs at the pit mouth [18], rather than with the maximum principal stress, which occurs at the base of the pit. These microcracks are often, but not always, associated with slip bands [17]. The TGSCC crack propagation mechanism has been widely accepted to follow Magnin's corrosion enhanced plasticity model (CEPM) [19]. This model explains the experimental observations of zig-zag type fractographic surfaces characteristic of boiling $MgCl_2$ induced SCC cracks in austenitic stainless steels [2,17,20]. Specifically, the model states that SCC typically grows along slip planes such that energy expenditure is minimized [19,21–23]. Nevertheless, since cracks are activated by stresses normal to that slip plane, they can grow along alternating parallel planes. For example, a crack along a {111} slip plane can zig-zag onto subsequent parallel {111} planes, resulting in overall crack growth in the ⟨112⟩ direction.

However, CEPM does not explicitly treat TGSCC propagation across grain boundaries. In general, the role of grain boundaries (GB) has received limited attention in TGSCC research. Jivkov and Marrow [24] proposed that the tensile elastic strain energy density of GBs should be considered in a model predicting the threshold stress intensity required for TGSCC crack propagation. However, they also suggested that this threshold stress intensity value has no physical basis since crack propagation rates decrease with decreasing stress intensity. Consequently, TGSCC propagation is often treated from a purely mechanical perspective. In fact, purely mechanical treatments such as Wenman's finite element approach [20] show close agreement with CEPM and experimental observations.

TGSCC has been shown to be driven by localized plastic deformation with environmental corrosion assistance [25,26]. Nevertheless, from a purely mechanical perspective, mechanical cleavage will occur when slip systems are activated by dislocation motion and pile-up against obstacles (e.g., Lomer locks [27] or grain boundaries). The criterion for slip system activation is represented by Schmid and Taylor factor. Schmid factor (m) represents the shear stress required to activate the easiest slip system, assuming isostress equilibrium [28], as shown in ref. [29]:

$$m = \frac{\tau_{CRSS}}{\sigma_y} \tag{1}$$

where $\sigma_y$ is the yield strength of the grain, and $\tau_{CRSS}$ is the critical resolved shear stress on the active slip plane. A high Schmid factor indicates that less stress is required to provide sufficient shear stress on the active slip system [29]. Grains with high Schmid factor are oriented such that easy slip is favorable, and slip steps can break through the passive film to initiate brittle cracking across a GB [30–32]. On the other hand, the Taylor factor (M) is derived based on the isostrain assumption in polycrystalline materials [28]. It is typically used to correlate macroscopic flow stress and the resolved shear stress, assuming multiple slip systems share the same shear stress and are thus equivalent to one another [33]:

$$M = \frac{d\gamma}{d\epsilon_x} = \frac{\sigma_x}{\tau} \tag{2}$$

where $\sigma_x$ is the macroscopic flow stress, $\epsilon_x$ is the grain strain, $\gamma$ is the shear strain on each slip plane, and $\tau$ is the shear stress on active slip planes.

The objective of this article is to understand how mechanical factors – namely, Schmid and Taylor factors – affect TGSCC behaviors at grain boundaries. Work will focus on a 304L austenitic stainless steel containing a central arc weld, representative of the weld seams on spent nuclear fuel dry storage containers believed to be susceptible to CISCC. Specimens are placed into four-point bend fixtures and undergo CISCC testing in a boiling $MgCl_2$ solution (following ASTM G-36). A large data set of transgranular crack behaviors in >300 grains along crack paths throughout the weld zone and heat-affected zone were collected using serial orientation imaging microscopy.

These experimental results enable the study of transgranular cracking behaviors with respect to the Schmid factor, Taylor factor, and their respective mismatches across grain boundaries.

## Results

### <u>**General Structure and Cracking Behavior**</u>

The microstructures of the BM, HAZ, and WZ are shown in the SEM-EBSD inverse pole figure (IPF) and image quality (IQ) figure, together with grain size and grain misorientation distributions in Figure 3. Both the as-received base metal and HAZ have uniform equiaxed grain structures containing a large amount of twinning. The average grain size of the HAZ ($27.7 \pm 21.7$ μm) is statistically equivalent to that of the BM ($29.8 \pm 17.6$ μm). Grain boundary misorientation distribution peaks at high angle range ($57.5°$ - $60.5°$) for both BM and HAZ, comprising ~50% of all boundaries in these regions. These high angle grain boundaries correspond to deformation twinning $\sum 3$ boundaries, as shown in Figure 3e and 3f.

As shown in Figure 3a and 3d, the weld zone is composed of large dendritic austenitic grains (>200 μm) mixed with δ-ferrite inter-dendritic phases (<15 μm). This structure is consistent with well-known post-weld cooling transformations, in which primary δ-ferrite develops from solidification at high temperature and then transforms to γ-austenite as temperature decreases [34]. Due to rapid cooling rates, the diffusion-controlled δ-γ transformation is incomplete, leaving islands of Cr-rich δ-ferrite retained within grains and near grain boundaries, as shown in Figure 2c [34–36].

To compare the SCC susceptibility across different zones, the number of cracks in the weld zone and HAZ are normalized by the respective length of these zones (because they share the same width), as shown in Figure 2d. A total of 138 cracks are found on the polished cross-section, with 31 of these cracks located in the weld zone and the remaining 107 cracks located in the HAZ. Note that no cracks are observed in the base metal, and henceforth the base metal will be excluded from the discussion. Most of the cracks are microcracks <10 μm in length, which exhibit crack initiation with limited propagation. A lower frequency of macrocracks >20 μm is present in both the weld and HAZ. From the normalized crack frequencies, the weld zone appears less susceptible to cracking than the HAZ, which is more susceptible than the base metal. Given the similar grain size and misorientation distribution between the HAZ and base metal, the greater cracking susceptibility of the HAZ may be due to its higher stress (since the HAZ is closer to the center of the coupon where 4-point bend stress is maximized) or sensitization [37–39].

Of the 37 cracks of interest selected for comprehensive crystallographic analysis by EBSD, 27 are located in the HAZ, and 10 are located in the WZ. Only 5 of these cracks initiate in an intergranular mode (IGSCC), and all of these are located in the HAZ, while the remaining 32 cracks initiate in a transgranular mode (TGSCC). IGSCC is widely accepted to be correlated with sensitization due to excessive heat input from thermal treatment or joining process, e.g. welding [37–39]. The dominance of TGSCC over IGSCC in this current study is attributed to the low carbon content in the UNS SS304L substrate used, which limits segregation to grain boundaries and sensitization during cooling [40,41]. Regardless of initiation type (i.e. IG or TG), all cracks in both the HAZ and WZ propagate transgranularly. Propagating TGSCC cracks show zig-zag deflection patterns with minor branching (Figure 2a), similar to what Alyousif and Nishimura reported at 428K (155°C) [42]. However, within the dendritic grains in the WZ, cracks also follow the Cr precipitated ferrite network. The reader is referred to our Data in Brief article [43] for comprehensive

SEM imaging and EBSD mapping of all 37 cracks of interest; these figures are provided only for representative cracking behaviors herein. Given the higher cracking susceptibility of the HAZ, the majority of the subsequent EBSD and EDS characterization focuses on the HAZ.

**Special Cracking Behaviors Revealed by EBSD**

The TGSCC crack propagation described in Section 3.1 exhibits some notable crack features, which will be highlighted here. Specifically, discontinuities and crack detours are observed in the HAZ on the scanned cross-section surface. In addition, crack terminations occur predominantly when crack tips arrive at grain boundaries (GBs). EBSD maps showing the Schmid factor and Taylor factor of all grains along crack paths containing representative crack detours, discontinuity, and terminations, are presented in Figure 4.

***Crack Detour***: Crack detouring describes a set of apparent discontinuities in the crack path around a hard grain – instead of propagating into the incident grain per the original crack path, the crack takes an alternative route around the grain, then resumes the original crack propagation direction on the opposite side of the grain. Crack detours are observed in 2 out of the 37 cracks examined by EBSD. Figure 4a is a representative example of a crack detour shown in SEM with its corresponding SEM-EBSD map. The cyan ($\overline{5}$ 5 $\overline{2}$) grain is the expected incident grain along the crack propagation path from the green (2 $\overline{15}$ 8) grain. However, as marked by the black crack path, the crack avoids the cyan ($\overline{5}$ 5 $\overline{2}$) grain. Instead, it takes an alternative route through the purple ($\overline{3}$ 5 3) grain before returning to the original crack path in the magenta (2 1 $\overline{1}$) grain. The cyan ($\overline{5}$ 5 $\overline{2}$) grain around which the crack detours has a lower Schmid factor (m=0.42) and higher

Taylor factor (M=3.29) than all of the surrounding grains (m ranges 0.47-0.5 and M ranges 2.37-2.88).

***Crack Discontinuity***: Only one crack discontinuity is observed among all 37 closely examined cracks, and it is shown in Figure 4b. This crack is relatively short (~16 μm) and has a straight and brittle cleavage appearance. The uncracked "discontinuity" is located at the intersection between the corner of the orange (1 27 4) grain and the major cyan (14 $\overline{5}$ $\overline{12}$) grain. This discontinuity occurs in the early-stage crack propagation in which the crack path is interrupted at a GB then immediately reappears on the same path on the other side of that grain. It is possible that this discontinuity is a jump of the crack across a grain, because no alternative crack direction is observed in the examined plane, nor is crack branching observed. Such a jump may be akin to the crack bridging phenomenon identified in IGSCC of austenitic stainless steels [44,45], which occurs when cracks deviate three-dimensionally around grains and along susceptible sensitized grain boundaries, leaving connected GB ligaments bridging behind the IGSCC crack tip [44,46]. However, the EBSD maps only show 2D views of the polished cross-section of the specimen, while SCC is 3D in nature [47]. Thus, this crack discontinuity could simply be a form of a crack detour out of the 2D examination plane [48], with the crack possibly being continuously connected in 3D.

***Crack Termination at Grain Boundaries***: Cracks tend to terminate exactly at GBs, rather than in the middle of a grain. Of the 37 grains selected for EBSD, 8 cracks are clearly identified to terminate at a GB (although due to the EBSD resolution limit, the exact endpoints of other crack tips are unidentifiable). These crack terminations at GBs mainly occur when the crack is propagating from a soft to harder grain. The large fraction of crack termini at GBs suggests that CISCC transmission across a GB is controlled by a competition between mechanical, chemical, and crystallographic factors. A representative crack tip terminating at a GB is shown in Figure 4c. It can be observed that after passing through the blue (13 $\overline{17}$ $\overline{17}$) grain (m=0.47, M=0.33), the crack tip stops at a low-angle grain boundary before the harder pink (3 12 4) grain (m=0.45, M=3.61).

## **Statistical Analysis of Schmid Factor and Taylor Factor**

Based on the EBSD maps of the 37 closely examined cracks, the Schmid factor and Taylor factor distributions of 338 grains are presented in Figure 5 and Figure 6, respectively. These distributions are given for (a) all <u>uncracked</u> grains on the specimen surface, (b) grains in which crack <u>initiation</u> occurred, and (c) all <u>propagation</u> grains along the 37 crack paths. Comparing distributions (a) and (b) informs preferential crack initiation orientations while comparing distributions (a) and (c) informs preferential crack propagation orientations.

From Figure 5a and Figure 6a, most of the grains on the specimen surface (i.e., the surface that had been in direct contact with the boiling $MgCl_2$) have a high Schmid factor (> 0.42) and high Taylor factor ($\geq$ 3.1), although these trends are more apparent in the WZ than in the HAZ. In order to define a quantitative criterion for comparing Schmid and Taylor factor distributions, the skewness of the distributions in Figure 5 and Figure 6 are calculated using Pearson's moment coefficient of skewness method [49] as:

$$\gamma = \frac{E[x^3] - 3\mu\sigma_{SD}^2 - \mu^3}{\sigma_{SD}^3} \tag{4}$$

where $\gamma$ is the sample skewness, $x$ is the data value, $E[x^3]$ is the third raw moment of x, $\mu$ is the mean, and $\sigma_{SD}$ is the standard deviation. From Figure 5, the Schmid factors in the WZ have dissimilar skewness between uncracked ($\gamma$ = -1.54) and initiation ($\gamma$ = -0.78) grains. However, Schmid factors in the HAZ have similar values of skewness for uncracked ($\gamma$ = -1.02), initiation ($\gamma$ = -0.78), and propagation ($\gamma$ = -1.07) grains. Likewise, Figure 6 shows that Taylor factors in the WZ also have dissimilar skewness between uncracked ($\gamma$ = -1.33) and initiation ($\gamma$ = -0.77) grains, while the skewness of the Taylor factor in the HAZ is consistent across uncracked ($\gamma$ = -0.44), initiation ($\gamma$ = -0.50), and propagation ($\gamma$ = -0.01) grains. The consistency of the skewness values in the HAZ demonstrates that the distributions of the Schmid factor or Taylor factor are statistically similar for all three scenarios. Thus, it can be concluded that crack initiation or propagation is not significantly affected by Schmid factor or Taylor factor alone.

**Analysis of GB character on cracking behavior**

It has been reported that special CSL GBs are generally IGSCC resistant in low SFE FCC metals such as stainless steels [50–52]. Thus, the effect of GB type and character on TGCISCC susceptibility is determined by comparing the occurrence of CSL GBs within the bulk HAZ to their occurrence along the crack paths, as shown in Figure 7. High-angle boundaries (HAB) are the predominant GB type in the bulk HAZ, while only 5.3% are low-angle boundaries (LAB). Coherent twin $\sum$3 boundaries are observed to be the major CSL GBs in the bulk HAZ, which is to be expected in stainless steel 304L due to their relatively low stacking fault energy (SFE) [53]. Twinning variants, e.g., $\sum$9 boundaries [50], comprise 6% of the CSL GBs in the bulk HAZ, and other CSL boundaries less than 2%.

Among all of the GBs associated with the 338 grains on crack paths, most of the crack propagations occur through HABs and $\sum 3$ twin boundaries. Meanwhile, a total of 38 notable discontinuous cracking behaviors (jump, detour and final termination as shown in Figure 4) are collectively classified as the "arrest at GB" case in Figure 7. Similar to crack propagations, the majority of these crack arrests occur at HABs and $\sum 3$ boundaries. Only two of the crack arrest cases occur at LABs, and only one each at $\sum 15$ and $\sum 27$ boundaries. Thus, the dominance of HAB and $\sum 3$ boundaries in both the propagation and arrest cases along the crack path is consistent with the overall distribution of HAB and $\sum 3$ boundaries in the bulk HAZ. Hence, GB character does not appear to significantly affect whether the crack will propagate, experience discontinuity, detour, or terminate at the GB. This result can be strengthened with additional data to provide a larger statistical sample, as well as by considering how crystallographic TGSCC cracks renucleate in each grain on different planes having different tilt and twist.

### FEA Stress vs. Crack Length

Initial FEA simulation results are summarized in Figure 8. Figure 8b reports the maximum von Mises stress near a crack tip of varying length, showing that stress concentration follows a typical parabolic shape of linear elastic fracture mechanics (LEFM) analysis [54]. This correlation aligns with the von Mises equivalent stress calculation around a micro-crack under uniaxial tension [55]:

$$\sigma_{Mises} = \frac{K_{IC}}{\sqrt{2\pi a}} \tag{5}$$

Where $K_{IC}$ is plane-stress fracture toughness and $a$ is crack length. This FEA result suggests that due to mechanical driving forces alone (i.e., neglecting effects of the corrosive environment), longer cracks experience a greater tendency to propagate. Figure 8c depicts the FEA stress distribution near a crack tip.

It should be noted that the FEA model in this study is constructed with the isotropic homogenous assumption and only considers elasticity. Future improvement of the simulation should implement crystal anisotropy and crystal plasticity to assess local stresses more accurately. However, the purpose of the current model is to provide relative values of von Mises and shear stress at cracks of differing lengths [18,48]. Since only plane stress is of interest in a static simulation scenario, pure elasticity analysis provides sufficiently accurate information for this purpose [56].

## Discussion

Schmid factor (m) and Taylor factor (M) of individual grain alone are not found to have significant effect on crack propagation. Thus, the mismatch between m and M of all grain pairs along the crack path of all 37 EBSD-analyzed cracks are investigated and recorded in Figure 9. These mismatch values represent the difference between $m_i$ and $m_{i-1}$ or between $M_i$ and $M_{i-1}$ in adjacent grains along the crack path from grain *i-1* to grain *i*. Grains with low m and high M behave as "hard" grains because they are generally more resistant to plastic deformation and exhibit more strain hardening [29,57]. On the contrary, "soft" grains are characterized by high m and low M. Both Schmid factor and Taylor factor describe plastic deformation in the grain relative to the applied stress. Low Taylor factor indicates that less shear strain is required in grain to accommodate the overall tensile strain [33]. Thus, reduced dislocation pile-up and lower localized strain occur at the grain boundary between a cracked grain and the low Taylor factor grain into which the crack is propagating [58–60]. On the other hand, grains with a high Taylor factor experience higher stress than their surroundings, leading to greater strain hardening and the development of a higher dislocation density to maintain strain compatibility with the neighboring grains [29,61,62].

TGCISCC propagation grain pairs concentrate in the second and fourth quadrants, indicating a "hard ↔ soft" grain pair preference. That is, TGCISCC propagation is most likely to occur between grains with opposite Schmid factor and Taylor factor mismatches. Cracks located in the second quadrant propagate from a high m, low M grain (soft) to a low m, high M grain (hard). By contrast, cracks in the fourth quadrant propagate from a low m, high M grain (hard) to a high m, low M grain (soft). Notable behaviors such as crack detouring, discontinuity, and termination at GBs are concentrated in the second quadrant (soft → hard), i.e., when a crack propagating in a softer grain meets a harder grain (lower m and higher M). Specifically, it should be noted that the crack detour in Figure 4a occurs in the neighboring soft ($\overline{3}$ 5 3) grain, while the other surrounding grains are hard grains with lower m and higher M. Since SCC usually branches and kinks in 3D [47], the "soft → hard" crack detour observations indicate that the crack is likely to be drawn towards and detours in the soft neighboring grain in 3D, rather than propagate directly through it. Crack discontinuity occurs between two grains with a large (> 1.0) Taylor factor difference, but the Schmid factor of the grains is similar.

The crack propagation mechanisms across GBs differ for "hard → soft" (fourth quadrant) propagation directions as opposed to "soft → hard" (second quadrant) propagation directions. In the "hard → soft" scenario, the low Taylor factor in the softer grain indicates higher shear stress on the slip plane, while a higher Schmid factor indicates less stress is required to activate slip systems [29,33]. Additionally, FEA simulations suggest that crack termination is always correlated with insufficient shear stress in the incident hard grain. Thus, when a crack tip approaches a softer grain, that softer grain deforms via slip, causing fracture across the GB to facilitate continuous crack growth [19,27,32].

With respect to the "soft → hard" scenario (second quadrant), to understand special cracking behaviors (discontinuity, detour, termination at GB), we consider three subgroups of grain pairs with similar Schmid and Taylor factor mismatch values; each encircled in Figure 9. Each of these grain pairs represents a crack-GB interaction. Since each of these crack-GB interactions in a subgroup has a similar Schmid and Taylor factor mismatch, any differences in their behavior (crack propagation vs. detour vs. termination) may be ascribed to differences in mechanical driving force associated with the length and position of the specific crack-GB interaction. A unique FEA simulation is executed for each crack-GB interaction in the selected subgroups, summarized in Table 2. The specific crack length and position used within these FEA simulations corresponds to the location of the crack-GB interaction from EBSD. The FEA calculated stress field at the crack tip indicates the localized tensile stress experienced by the grain on the opposite side of the GB upon which the crack is incident.

The FEA resultant von Mises stress at the crack tip is converted to shear stress using both the Schmid factor and Taylor factor, according to Equations 1 and 2. These shear stresses are plotted in Figure 10 for all three subgroups of data, and the instances of crack propagation, termination, and detouring are distinguished symbolically. These plots show that cracks terminate at the GB when shear stress is lowest, while crack propagation and detouring occur at higher shear stress. This demonstrates that crack propagation through a GB requires a threshold level of shear stress (although this threshold between crack termination and propagation differs for each subgroup based on the Schmid and Taylor mismatches). When the shear stress is below this threshold, cracks will terminate due to insufficient plastic deformation. However, there is no clear shear stress dependence to discern when cracks will propagate or detour. This can be explained by two factors. First, mechanism 3 requires crack propagation in adjacent grains to be more favorable than that in the directly incident grain – this requires knowledge of the crystallographic orientation of all adjacent grains, which EBSD inherently cannot provide. Second, detouring tends to occur in 3D [47], while the FE simulation uses a 2D plane stress condition. In addition, prior experimental observations of intergranular crack bridging by Babout et al. [44] and Marrow et al. [45] are associated with special CSL grain boundaries; our future work will expand the analysis herein to consider the role of grain boundary type and misorientation.

The FEA results of similar m and M subgroups in Figure 10 suggest a possible threshold level of shear stress required for the crack to propagate through a GB (although this threshold differs for each m-M subgroup based on the mismatch values). Similar stress thresholds are also reported in past literature [19,63,64], but no agreement on the specific value has been achieved yet.

There remain outliers in the mismatch map (Figure 9), which indicates that local strain and stress incompatibility, as represented by the Schmid factor and Taylor factor, may not be sufficient to comprehensively describe and predict TGCISCC behaviors. Thus, other factors, such as temperature [42,65,66], chloride concentration [67,68], and pH [5,66,69], should also be considered when evaluating TGCISCC phenomena. Examination of additional cracks would also improve the statistical significance of trends identified herein.

## Methods

### **Material, Weld, and Residual Stress**

This work utilized the ASTM G39 bent-beam method for evaluating the stress-corrosion behavior of alloy sheets [70]. Two hot rolled and pickled commercial SS 304L (UNS S30403) sheets of 3 mm thickness with the composition shown in Table 1 were used to prepare the specimen coupons. To create specimens most representative of the vertical seam weld in spent nuclear fuel (SNF) canisters based on the Sandia National Laboratories' Canister Mockup Report [71], 30-degree bevels were prepared on sheet edges. Two-pass gas tungsten arc welding (GTAW) was performed to join the sheets along the beveled edges, using SS 308L (UNS S30880) filler. GTAW parameters were 110 A current and 12 V voltage; the interpass temperature was 23.9 ºC during the first pass and 130 ºC during the second pass. The welded sheet was laser cut into coupons of dimensions 105 mm × 18.5 mm × 3 mm, with the welding seam oriented laterally and centered on the axial length of the coupon, as shown in Figure 1. Subsequently, the weld bead was ground flat in the transverse direction (TD) with commercial 100 grit grinding paper on a surface grinding machine.

The average residual tensile stress at longitudinal welds in SNF canisters was 380 MPa, based on deep hole drilling measurements [71]. To emulate this tensile stress for SCC initiation in the laser-cut coupons, the coupons were loaded into a four-point bend fixture made from Hastelloy C-276. Bending stress was applied with a loading bolt [43]. To calculate the applied stress, the coupon deflection was determined by a strain gauge. The maximum tensile stress applied on the convex side of the coupon was calculated based on the equation [70]:

$$\sigma = \frac{12Et}{(3H^2 - 4A^2)} \tag{3}$$

where $\sigma$ is maximum tensile stress, $E$ is the modulus of elasticity, $t$ is the thickness of specimen, $y$ is maximum deflection (between outer supports), $H$ is the distance between outer supports (3.5 inches), and $A$ is the distance between inner and outer supports (0.875 inches). It is worth noting that Equation 3 is only valid in the linear elastic regime. But the four-point bending introduced plastic deformation, so the actual maximum stress on the specimen was likely lower than the target 380 MPa (this was subsequently corroborated by finite element analysis, presented later in this section). After the coupon was loaded into the four-point bend fixture, it was set in the air for at least a few days before performing corrosion tests for repassivation.

## Boiling Magnesium Chloride Corrosion Test

Boiling magnesium chloride ($MgCl_2$) corrosion testing was performed in the Advanced Materials Laboratory at the University of New Mexico and Sandia National Laboratories; the setup is shown in accompanying paper [43]. The boiling solution was made by adding deionized (DI) water to the reagent-grade magnesium chloride hexahydrate flake in the boiling flask. The bent coupon was fully immersed in the boiling $MgCl_2$ solution while loaded in the bending fixture. To maintain the required 155 ± 1°C (311.0 ± 1.8°F) boiling temperature, the concentration of the $MgCl_2$ solution was set to 54.3 wt% [72]. Cooled DI water cycled through the condenser to capture the water vapor generated from the boiling of the $MgCl_2$ solution and maintain a stable boiling temperature.

## Metallurgical Preparation and Electron Microscopy Characterization

Cross-sections taken on the longitudinal direction and normal direction (LD-ND) plane from the pre-exposed coupons were prepared for mechanical and microscopic examination. An area of interest that spanned the weld zone (WZ), heat-affected zone (HAZ), and base metal (BM) was sectioned using a diamond circular saw to minimize the introduction of deformation and residual stress, as outlined in Figure 1. For optimal edge retention, sectioned specimens were mounted in graphite-blend conductive resin with the cross-section facing out; the press mounting was conducted at 200 °C, 3000 psi for 12 minutes. Mechanical polishing was conducted using a Buehler EcoMet30 semi-automated polishing system using 360, 600, 800, 1000, and 1200 grit SiC paper, with 20 N force and 200 rotations per minute (rpm) for 1-2 mins at each grit, followed by 6 μm, 3μm, and 1μm diamond suspension with 15 N force and 100 rotations per minute (rpm) for 3 mins each. Final mirror-like polishing was achieved through vibratory polishing with 0.05 μm colloidal silica for 4-6 hours. After vibratory polishing, samples were immediately transferred into a micro-organic soap solution and ultrasonically cleaned for 15 minutes to remove residual silica. Final ultrasonic cleaning with acetone and isopropanol was followed by air-blow drying.

Cracks were identified across the cross-sectioned specimen using a ThermoFisher Scientific Helios G4 UX Dual Beam Scanning Electron Microscope (SEM). SEM imaging conducted at 5 kV, 0.4 nA was used to count the number of cracks and measure crack lengths across the BM, HAZ, and WZ. HAZ is identified as the region with the color band produced from surface oxidation during welding, as shown in Figure 1a. Figure 2a and 2b are the representative crack morphologies and grain structure in the HAZ and WZ revealed by ion polishing.

Crystallographic and compositional characterization was completed with a Quanta 650 FEG SEM equipped with an EDAX Hikari™ electron backscatter diffraction (EBSD) detector and an Oxford Aztec Xstream-2 silicon drift Energy-dispersive X-ray spectroscopy (EDS) detector. A total of 37 cracks, including all major cracks that are longer than 20 μm and some shorter cracks (<20 μm) that contained features of interest (e.g., straight cleavage, discontinuity), were selected for comprehensive crystallographic analysis. Grain orientations along the entire length (i.e., from crack mouth to tip) of these 37 cracks of interest were subsequently mapped using serial EBSD.

The voltage of the SEM EBSD-EDS scanning was set at 20 kV, and the spot size was 5.5. Dwell time and step size were 1 ms and 0.3-2.5 μm/pixel, respectively. The neighboring confidence index (CI) was set to >0.1, and the neighboring grain tolerance angle was 3°. The grain orientation, Schmid factor, and Taylor factor of 338 grains and corresponding GBs of interest were extracted and analyzed using the EDAX OIM 8 software. Low-angle boundaries (LAB) (<15°) and high-angle boundaries (HAB) (15° - 180°) were marked, together with $\sum 3$ to $\sum 11$ coincidence site lattice (CSL) [73] grain boundaries in distinct colors, as shown in Figure 3.

Note that because of preferential polishing at the interface between the crack initiation surface and the mounting epoxy, the edge of the corroded specimen could not be indexed by the EBSD detector. This caused a loss of EBSD information over the first 5-10 μm of the crack initiation surface, but it did not hinder the identification of crack initiation mode since the grain sizes in all zones (>20 μm) exceeds this un-indexable region.

**<u>Finite Element Analysis (FEA)</u>**

To analyze the mechanically-driven stress concentrations near crack tips of varying configurations, Abaqus 2020 was used to simulate the four-point bend test. For simplicity, a plane-stress rectangular specimen was created with $L \times H = 105.5 \times 3.5 \ mm$, and Young's modulus and Poisson's ratio of the entire specimen were assumed to be that of stainless steel 304L at 190 GPa and 0.27, respectively. As shown in Supplementary Figure S1, the bottom of the specimen was pinned symmetrically at horizontal positions $x$= -44.5 and 44.5 mm, with displacement $u_x =$ $u_y = 0$. The displacement was applied at positions $x$= -22.2 and 22.2 mm with a magnitude equal to -1.016 mm.

An individual, one-dimensional crack was oriented along the $y$-direction in the simulation. A series of simulations were conducted, with the crack lengths and horizontal ($x$-direction) positions of the crack informed by EBSD maps, as summarized in Table 2. The width of the simulated cracks were calculated from their experimentally measured lengths by assuming a crack aspect ratio of 20 (length/width), following Musienko's work [74].

The mesh was created using CPS8 [75], which is an 8-node biquadratic plane stress quadratic element. After a convergence study, the mesh size near the tip was set to be 0.01 mm, and in the region further from the tip, the mesh size was set to be 0.02 mm. The global seed was selected to be 0.1 mm. The mesh near the tip is shown in Supplementary Figure S2. The von Mises stress and normal stress $\sigma_{xx}$ were used in the analysis to compare the stress at crack tips.

**Data Availability:**

The original EBSD datasets and corresponding images that support the findings of this study are available at: doi: 10.17632/23wxhgd55p.5. A detailed explanation of data processing is published as a separate submission to Data in Brief at: https://doi.org/10.1016/j.dib.2022.108059.

**Acknowledgments:** The authors are grateful to Dr. Wenbin Yu of the School of Aeronautics and Astronautics at Purdue University for his supervision and funding support for F. Tao. The authors also acknowledge Dr. Talukder Alam at Purdue University for assistance with SEM-EBSD. Sandia National Laboratories is a multi-mission laboratory managed and operated by National Technology and Engineering Solutions of Sandia, LLC., a wholly owned subsidiary of Honeywell International, Inc., for the U.S. Department of Energy's National Nuclear Security Administration under contract DE-NA0003525. This paper describes objective technical results and analysis. Any subjective views or opinions that might be expressed in the paper do not necessarily represent the views of the U.S. Department of Energy or the United States Government. Financial assistance from the U.S. Department of Energy's Nuclear Energy University Program under contract DE-NE0008759 is also acknowledged. This document is SAND2022-5221 J.

**Author Contributions:**

H.J.Q. and J.P.W. conceived the study. H.J.Q., T.M., and J.M.T. performed boiling $MgCl_2$ test, F.T. and N.G. performed the FEA analysis. H.J.Q. and F.T. analyzed the data and drafted the original manuscript. R.F.S., E.J.Q., and J.P.W. supervised the project and revised the manuscript. E.J.Q. and J.P.W. acquired the funding. All authors reviewed and approved the final manuscript.

**Competing Interests:**

The authors declare no competing interests.

Figure Captions

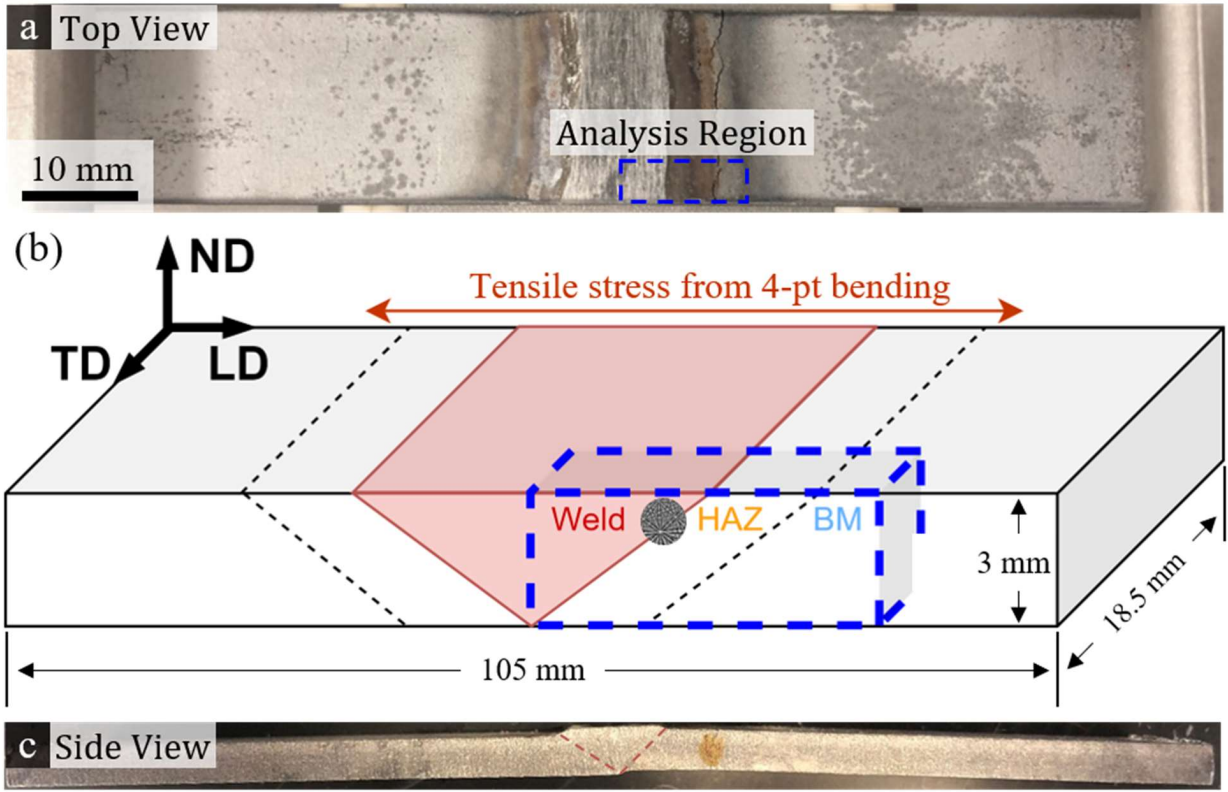

Figure 1. Welded coupon after stress corrosion crack testing in boiling magnesium chloride solution.

(a)(c) Top view and side view; (b) schematic illustration of the welded sample dimension, three different zones produced by GTA welding, and final EBSD analysis regions, marked by Kikuchi pattern. (Images are adapted from ref. [43])

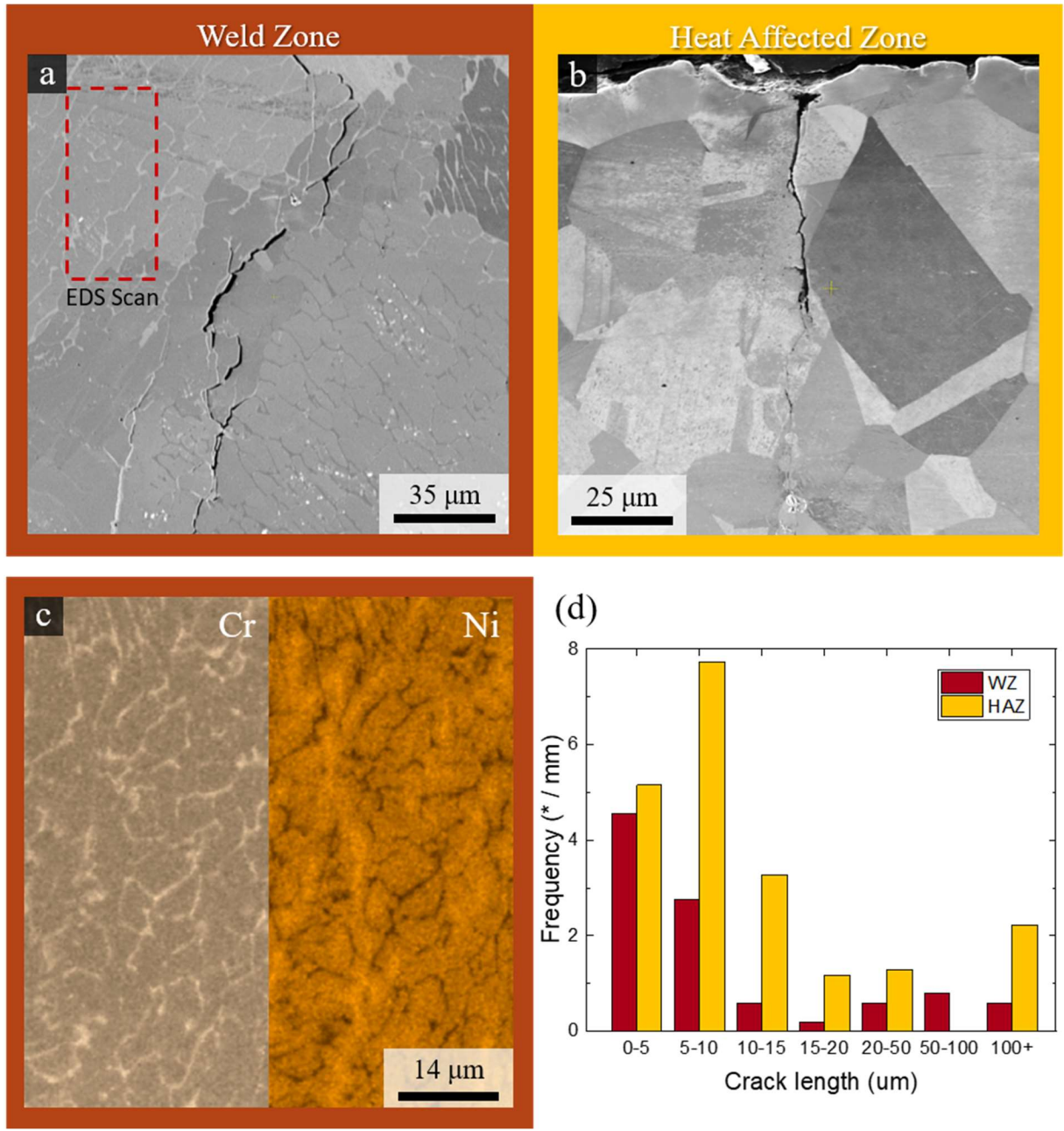

Figure 2. Crack morphology.

(a) HAZ and (b) WZ; (c) EDS map of WZ shows precipitation along subgrain boundaries: Ni depletes while Cr enriches at the subgrain GBs; (d) Crack length distribution in HAZ and WZ on the ND-LD plane of the sectioned coupon.

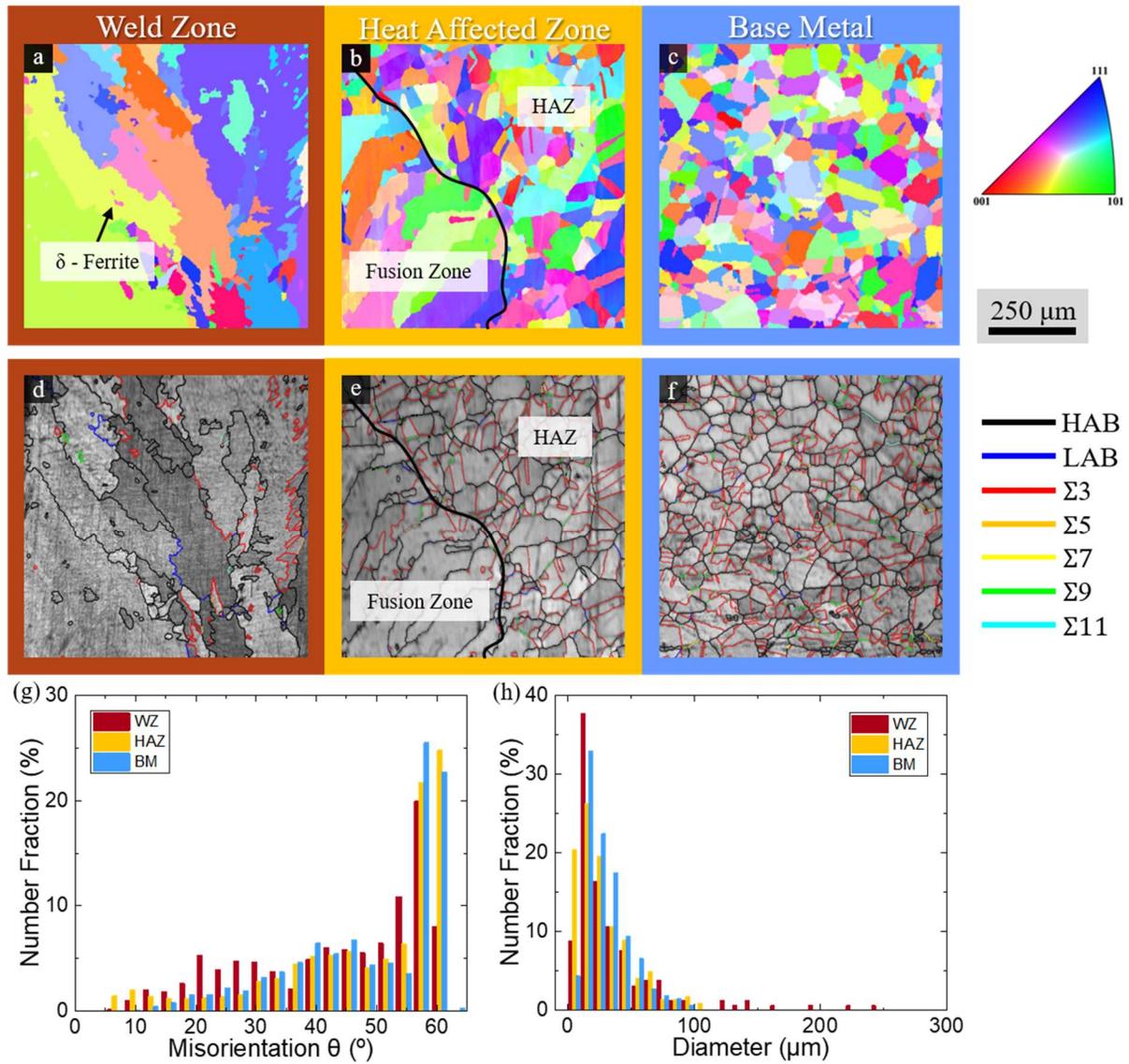

Figure 3. General crystal structure of base metal (BM), heat affected zone (HAZ), and weld zone

(WZ) in as-received specimen revealed by EBSD.

(a)(b)(c) Inverse pole figures (IPF); (d)(e)(f) image quality figures (IQ); (g) grain size distribution;

(h) grain boundary misorientation angle distribution.

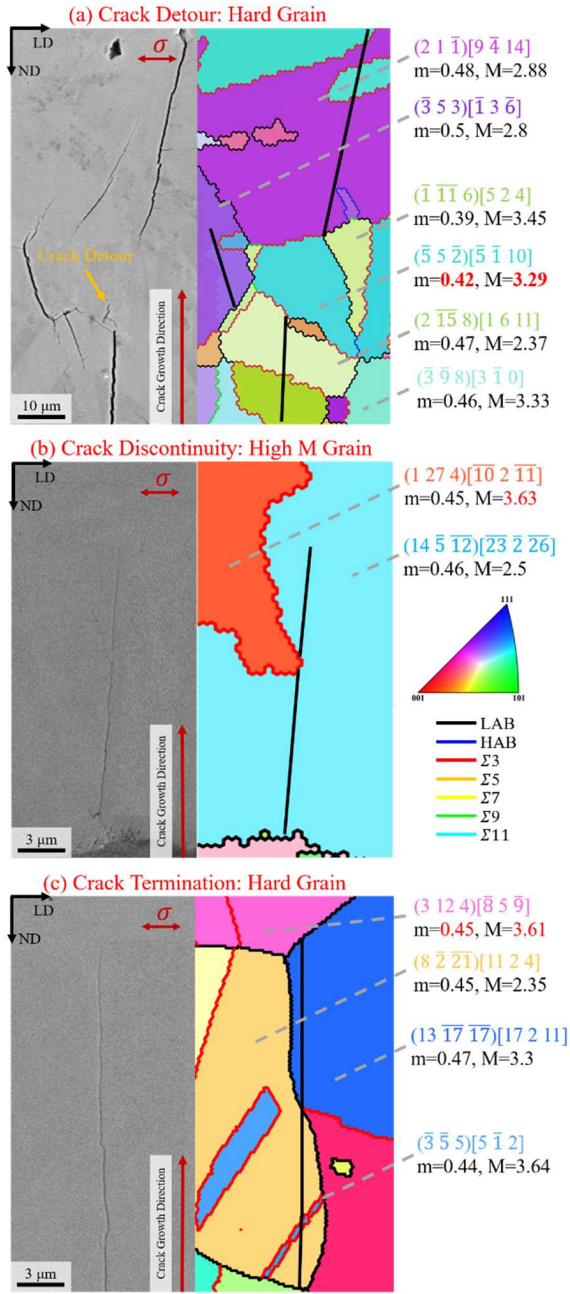

Figure 4. SEM morphology and EBSD maps of representative crack paths for different cracking behaviors.

(a) crack detour; (b) crack discontinuity; (c) crack termination at GB. All Schmid and Taylor factors are identified in all grains along the crack paths. (Images are adapted from ref. [76])

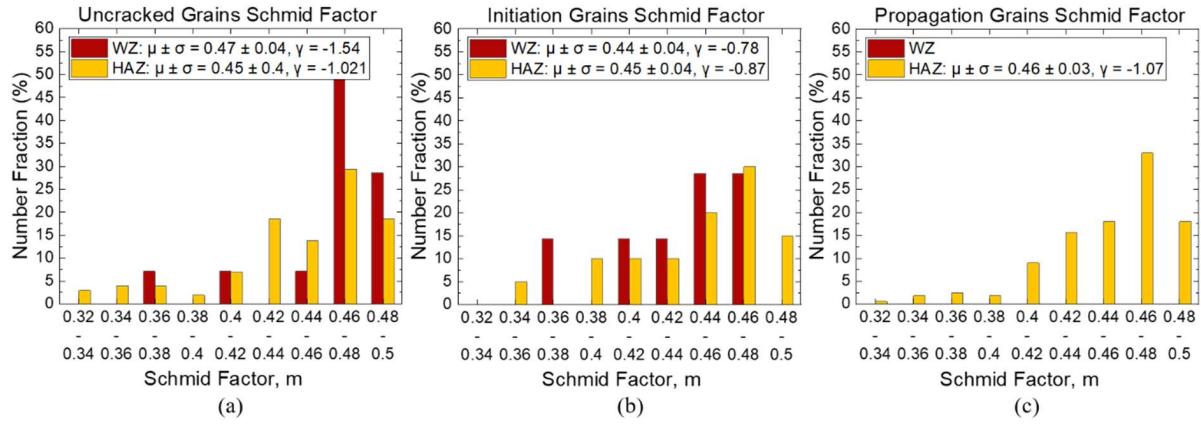

Figure 5. Schmid factor distributions.

(a) uncracked surface grains; (b) surface initiation grains; (c) propagation grains along crack path.

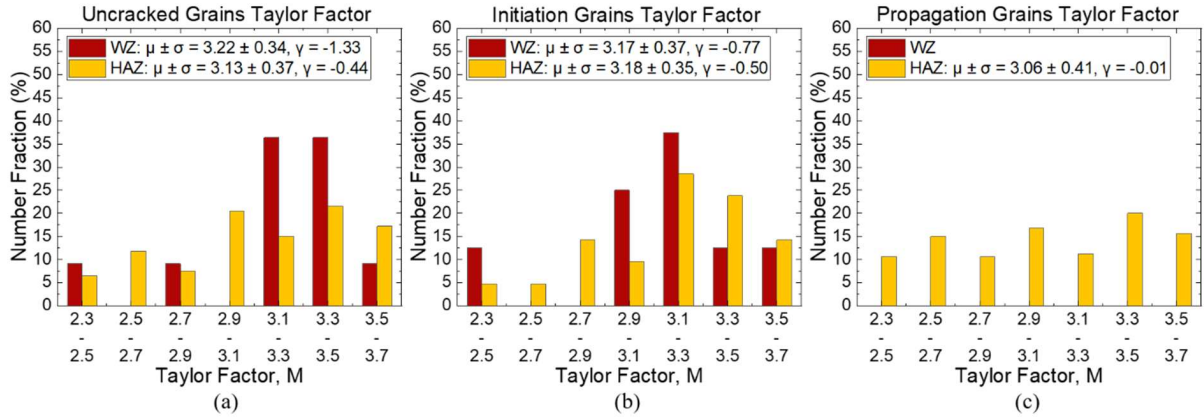

Figure 6. Taylor factor distributions.

(a) uncracked surface grains; (b) surface initiation grains; (c) propagation grains along crack path.

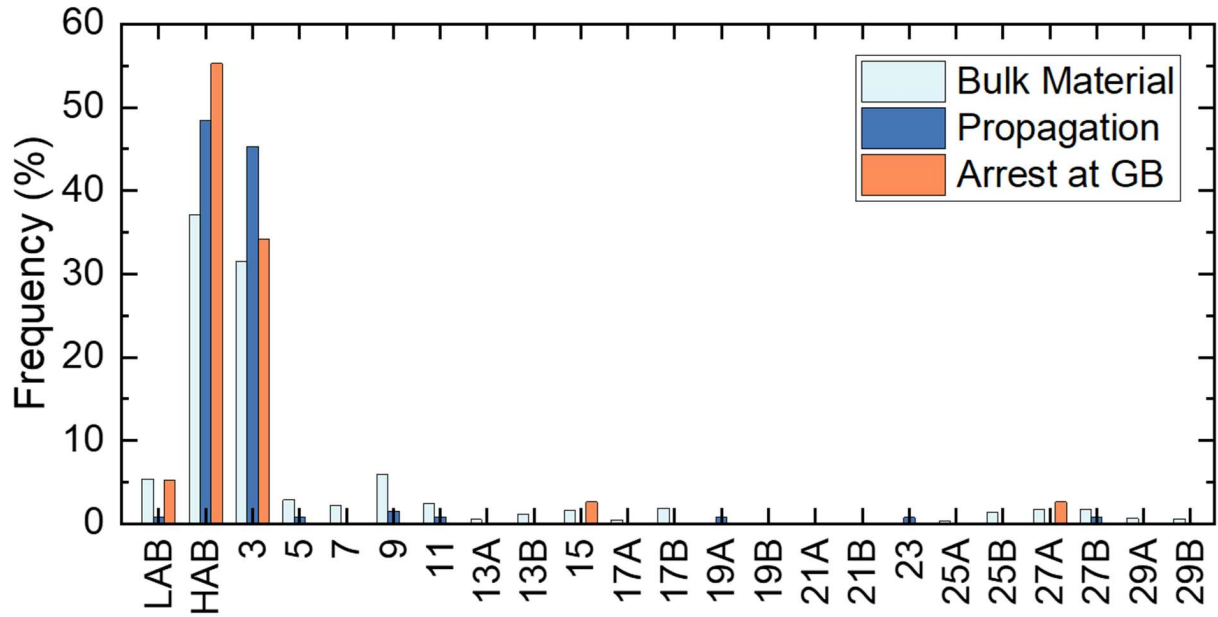

Figure 7. The GB type and CSL boundaries distribution of the bulk material, propagation and arrest scenarios along the 37 crack paths.

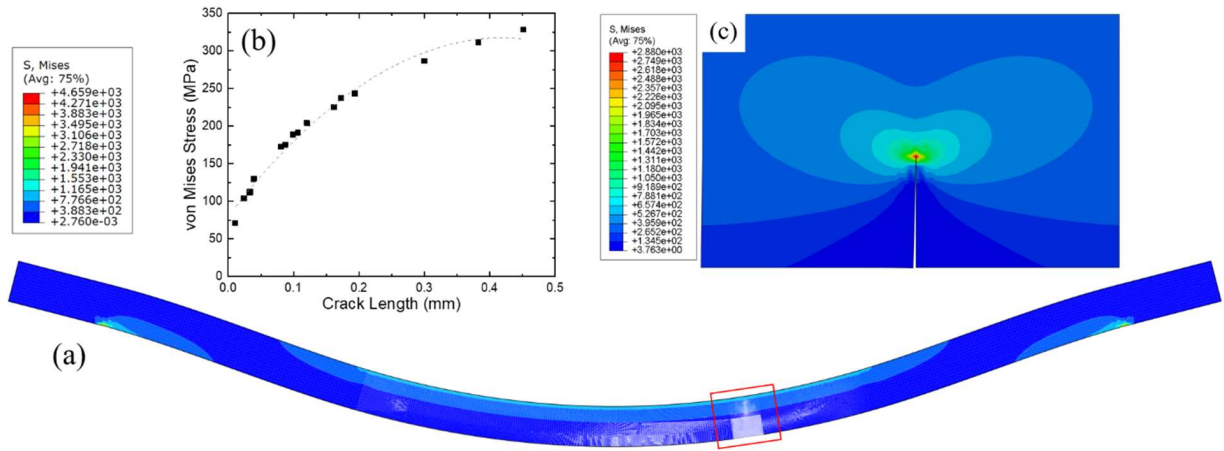

Figure 8. FEA results.

(a) Deformed 4-point bent specimen; (b) von Mises stress at different crack lengths for all cracked

grain pairs; (c) high magnification view of the stress distribution near the crack tip.

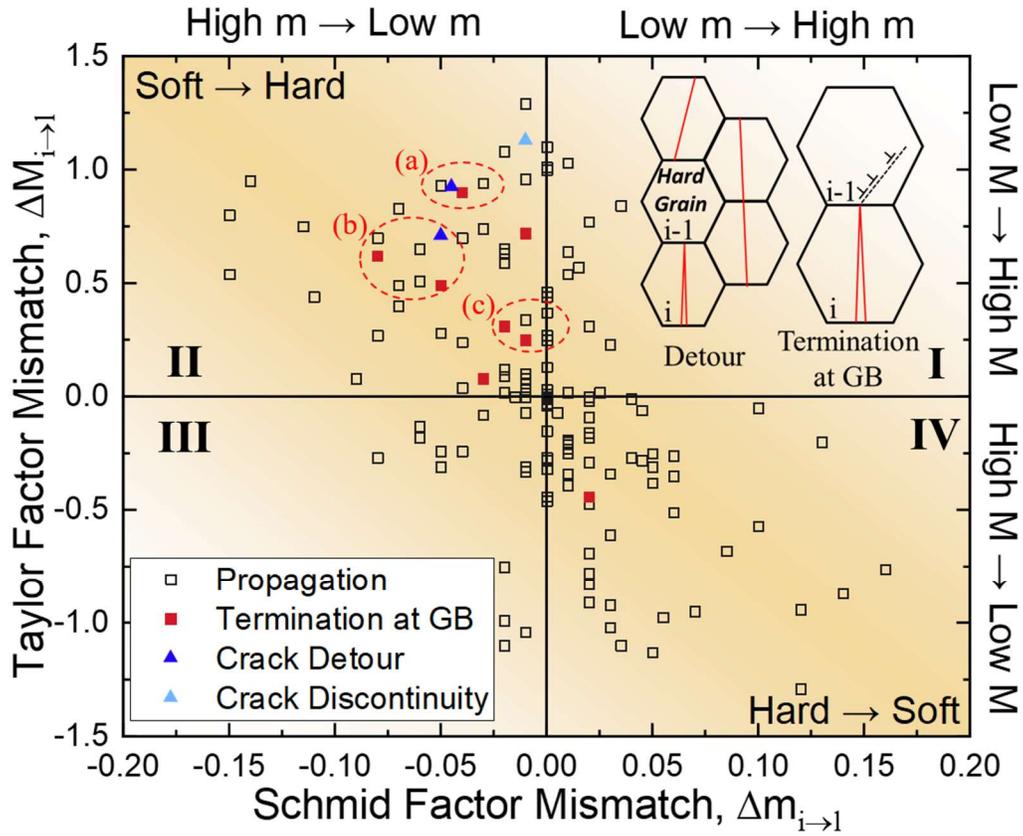

Figure 9. Schmid-Taylor factor mismatch map for transgranular SCC propagation paths.

Each data point represents a grain pair (i.e., GB-crack interaction) along the crack path; data are gathered from all 37 EBSD-analyzed cracks. Subgroups of data points encircled (a)(b)(c) are further analyzed in

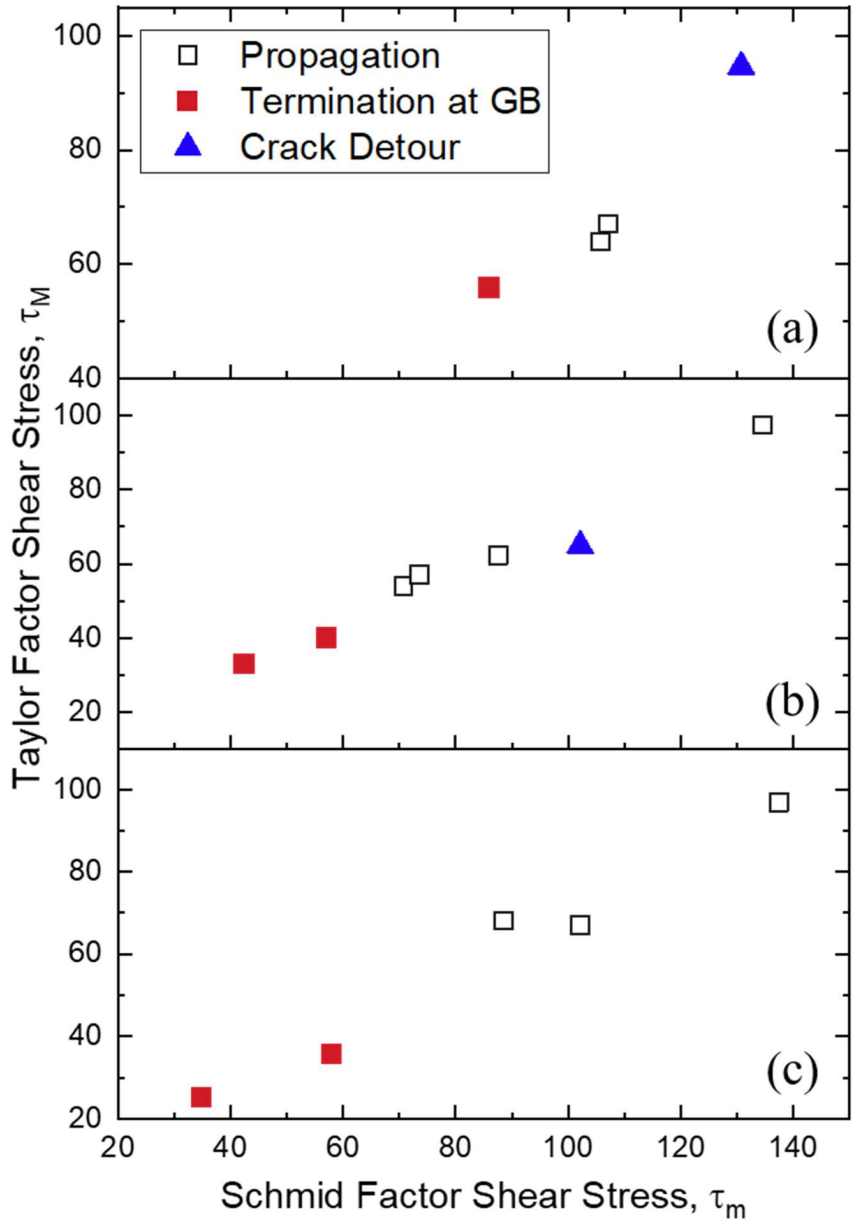

Figure 10.

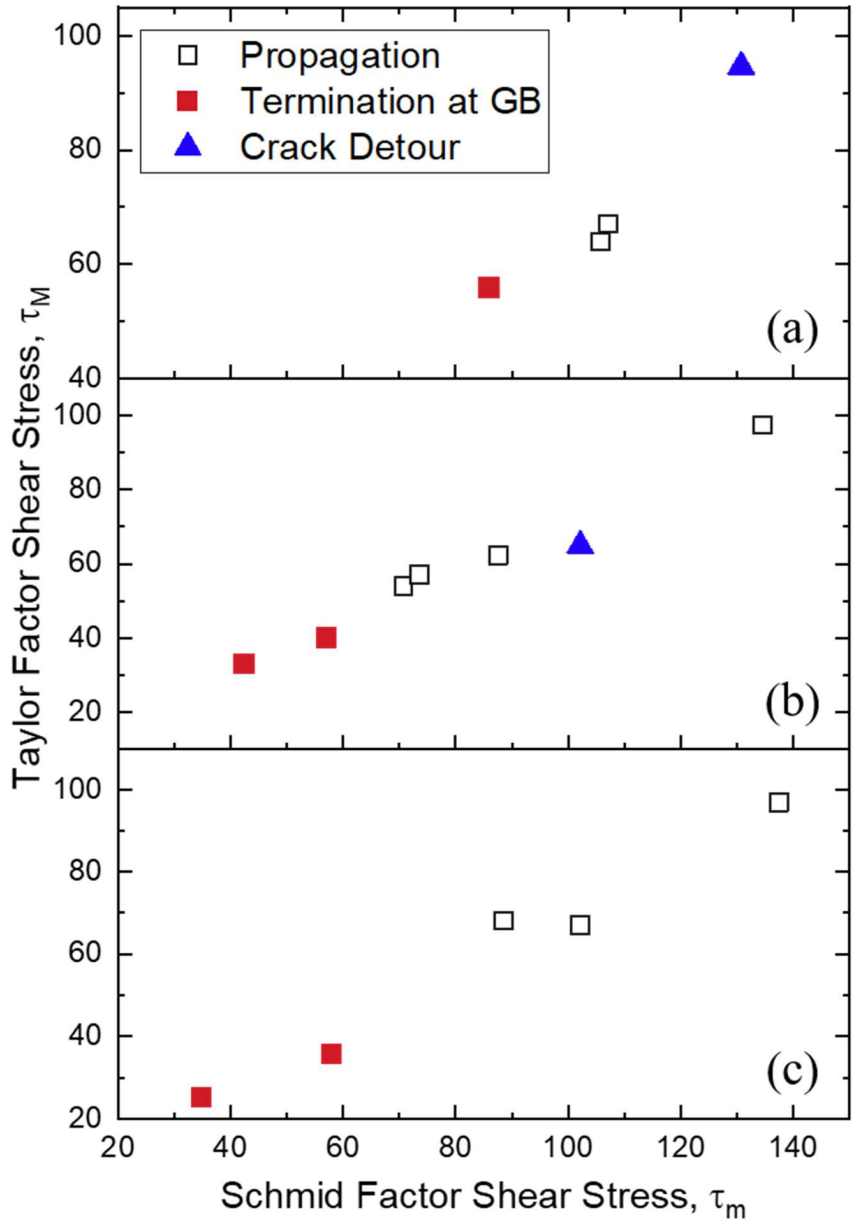

Figure 10. Schmid factor shear stress and Taylor factor shear stress of the grain pairs within subgroups (a - c) as identified on mismatch map (Figure 9).

*Table 1. Alloy and weld metal composition (wt.%).*

| Materials | Alloying wt.% balance Fe | | | | | | | | | |
|---|---|---|---|---|---|---|---|---|---|---|
| | C | Si | Cr | P | S | N | Mn | Ni | Mo | Cu |
| S30403 | 0.027 | 0.35 | 18.11 | 0.023 | 0.04 | 0.056 | 1.31 | 8.02 | - | - |
| S30880 | 0.014 | 0.47 | 19.88 | 0.021 | 0.002 | - | 1.83 | 9.66 | 0.01 | 0.1 |

*Table 2. Summary of FEA simulation results, organized by subgroup of crack-grain boundary interactions corresponding to encircled data points in Figure 9.*

| Grain Pair ID | Cracking Behavior | Length (mm) | Position from specimen center (mm) | Schmid Factor | Taylor Factor | Von Mises stress at crack tip (MPa) |
|---|---|---|---|---|---|---|
| A1 | Propagation | 0.162 | 11.6 | 0.47 | 3.52 | 225.1 |
| A2 | Propagation | 0.033 | 11.3 | 0.41 | 3.35 | 111.5 |
| A3 | Propagation | 0.194 | 9.3 | 0.44 | 3.63 | 243.5 |
| A4 | Termination at GB | 0.107 | 9.6 | 0.45 | 3.42 | 191.0 |
| A5 | Detour | 0.383 | 10.4 | 0.42 | 3.29 | 311.4 |
| B1 | Propagation | 0.081 | 10.3 | 0.41 | 3.2 | 172.5 |
| B2 | Propagation | 0.121 | 9.4 | 0.43 | 3.27 | 203.8 |
| B3 | Propagation | 0.087 | 11.3 | 0.42 | 3.07 | 175.1 |
| B4 | Propagation | 0.452 | 10.4 | 0.41 | 3.37 | 328.2 |
| B5 | Termination at GB | 0.039 | 10.3 | 0.44 | 3.25 | 129.8 |
| B6 | Termination at GB | 0.024 | 12.7 | 0.41 | 3.15 | 103.5 |
| B7 | Detour | 0.172 | 12.6 | 0.43 | 3.66 | 237.6 |
| C1 | Propagation | 0.100 | 11.6 | 0.47 | 2.77 | 188.6 |
| C2 | Propagation | 0.301 | 11.6 | 0.48 | 2.95 | 286.3 |
| C3 | Propagation | 0.120 | 10.3 | 0.5 | 3.05 | 204.3 |
| C4 | Propagation | 0.300 | 10.4 | 0.48 | 2.95 | 286.4 |

| C5 | Termination at GB | 0.039 | 8.5 | 0.45 | 3.61 | 128.9 |
| C6 | Termination at GB | 0.011 | 8.8 | 0.49 | 2.82 | 71.1 |